\documentclass[prb,twocolumn,nolongbibliography,floats,10pt,aps,superscriptaddress]{revtex4-2}
\usepackage{amsmath}
\usepackage{amsfonts}
\usepackage{braket}
\usepackage{xcolor}
\usepackage{appendix}
\usepackage{enumitem}
\usepackage{graphicx}
\usepackage{float}
\usepackage{tikz}
\usepackage{algorithm}
\usepackage{algpseudocode}
\usepackage{bbm}
\usepackage{mathtools}

\RequirePackage[hyperindex,colorlinks,bookmarksnumbered, plainpages=true,pdfstartview=FitH]{hyperref}
\hypersetup{linkcolor=blue,urlcolor=blue,citecolor=blue}
\usepackage{hyperref}

\begin{document}
  \title{Matrix product states and first quantization}
  \author{Jheng-Wei Li}
  \affiliation{Universit\'e Grenoble Alpes, CEA, Grenoble INP, IRIG, Pheliqs, F-38000 Grenoble, France}

  \author{Xavier Waintal}
  \affiliation{Universit\'e Grenoble Alpes, CEA, Grenoble INP, IRIG, Pheliqs, F-38000 Grenoble, France}

\begin{abstract}
Common wisdom says that the entanglement of fermionic systems can be low in the second quantization formalism but is extremely large in the first quantization.
Hence Matrix Product State (MPS) methods based on moderate entanglement have been overwhelmingly formulated in second quantization.
Here we introduce a first-quantized MPS approach to simulate quantum many-body systems.
We show that by reformulating the way the fermionic anti-symmetry is handled, we arrive at
MPS with a level of entanglement comparable to the usual one found in second quantization.
We demonstrate our scheme on the one-dimensional $t$-$V$ model (spinless fermions with nearest neighbour density-density interaction) for both ground state and time evolution. For time evolution, we find that the entanglement entropy in first quantization is significantly smaller than in its second quantization counterpart.
\end{abstract}
\date{\today}
\maketitle
  
\textit{Introduction.---}
Consider a system of $N$ particles and $L$ sites. Entanglement means two different things in the first quantization compared to the second. 

In second quantisation, a many-body wavefunction $\ket{\Psi}$ takes the form,
\begin{equation}
     \ket{\Psi} = \sum_{n_1, \ldots, n_L} \psi (n_1, \ldots, n_L) \ket{n_1, \ldots, n_L},
\end{equation} where $n_\ell \in\{0,1\}$ is the occupation number of single-particle state $\ell\in\{1,\ldots,L\}$ and $\sum_{\ell} n_{\ell} = N$. 
Entanglement is a statement about the rank of the $L$-legs tensor $\psi (n_1, \ldots, n_L)$, it measures how different \emph{regions} (set of sites) are interlinked. 
For many problems of interest, the bipartite entanglement is small, allowing $\psi$ to be represented accurately by a Matrix Product State (MPS) \cite{Vidal2003a, PerezGarcia2006, Hastings2007, Eisert2010, Brandao2014}. 
For short-ranged correlated systems in one dimension, the entanglement satisfies the so-called entropic area law making the MPS representation particularly successful.  
Variational methods based on MPS have become central techniques of the quantum many-body toolbox, solving large problems despite the exponentially large $\sim 2^L$ dimension of the Hilbert space.
A plethora of algorithms have been developed to study ground states, finite-temperature density matrices, real-time dynamics, and spectral functions for many-body systems \cite{Verstraete2008, Schollwoeck2011, Chan2011, Stoudenmire2012, Bridgeman2017, Paeckel2019a, Baiardi2019, Vanderstraeten2019, Orus2019}.
Almost, if not, all these techniques are formulated within the second quantization picture. 
 
In first quantization,  the $N$-particle state now reads
\begin{equation}
\ket{\Psi} = \sum_{x_1, \ldots, x_N} \psi (x_1, \ldots, x_N) \ket{x_1, \ldots, x_N},
\end{equation} 
where $x_i\in\{1,\ldots,L\}$ is the position (occupied site) of the $i^{th}$ particle and $\psi$ is now a $N$-legs tensor.
\emph{Particle}-entanglement is a statement about the rank of $\psi (x_1, \ldots, x_N)$, a very different object from its second quantization counterpart.
Particle-entanglement measures how a set of \emph{particles} is interlinked with another set. 
Crucially, for indistinguishable particles such as fermions, the wave function needs to be  properly anti-symmetrized with respect to particle exchange, 
\begin{equation}
\psi (\ldots, x_i, \ldots, x_j, \ldots) = -  \psi (\ldots, x_j, \ldots, x_i, \ldots)
\end{equation}
Pauli principle implies that fermions exhibit a large particle entanglement \cite{Coleman1963, Shi2003, Iblisdir2007, Haque2007, Zozulya2008, Amico2008, Haque2009, Killoran2014, Carlen2016, Barghathi2017,  Lemm2017, Radhakrishnan2023a}.
We measure the entanglement of the tensor $\psi$ using the von Neumann entropy $S_P = -\text{Tr}\ [\rho_P \log \rho_P]$ of the reduced density matrix 
$\rho_P = \text{Tr}_{\{ x_i,\ i>P\} } [\psi \otimes\psi^*]$ of a subset of $P$ particles 
($P$ sites in second quantization). 
It is generally believed that the particle entanglement entropy $S_P$ satisfies
$S_{P} \geq {\rm{ln}} \binom{N}{P}$ in sharp contrast to its second quantization counterpart. This is Coleman’s theorem for $P=1$,
a conjecture for $P=2$ \cite{Carlen2016} (with equality for a Slater determinant)
with numerical evidences for $P<4$.
Since the memory footprint of a MPS typically scales as $e^{S_{N/2}}\approx 2^N$,
i.e. is exponentially large,  it follows that using MPS in the first quantization framework is generally considered \emph{not} to be a good idea. 

In this work, we show that, by formulating the problem slightly differently, an efficient first quantization MPS approach can nevertheless be constructed for both ground state calculations and real-time dynamics.

\textit{Model.---} For concreteness, we consider the
well-known $t$-$V$ model:  spinless fermions in one dimension with nearest neighbour
hopping $t$, nearest neighbour density-density interaction $V$, and open boundary conditions. The Hamiltonian reads,
\begin{align}
 H_{t-V} = -t\sum_{\ell=1}^{L-1}(c^{\dagger}_{\ell} c_{\ell+1} + {\rm{h.c.}}) 
 +V \sum_{\ell=1}^{L-1} \hat{n}_{\ell}\hat{n}_{\ell+1},
\end{align}
where $c^{\dagger}_{\ell}$ and $c_{\ell}$ are fermionic creation and annihilation operators at site $\ell$ and $\hat{n}_{\ell} = c^{\dagger}_{\ell} c_{\ell}$ is the number operator. In first quantization, $H=H_t+H_V$ takes the form,
\begin{align}
  H = -t\sum_{n=1}^{N}D_{n} + V\sum_{n<m} v(x_{n}, x_{m}),
  %H = -\sum_{i}(c^{\dagger}_{i} c_{i+1} + h.c.) + V\sum_{i}n_{i}n_{i+1}
  \label{eq:hamQ1}
\end{align}
with $D_{n}\psi = \sum_\pm \psi(\ldots,x_{n-1},x_n\pm 1,x_{n+1},\ldots)$
(in the continuum, one would have $D_{n} = \partial^2/\partial x_n^2$)
and $v(x,x')= \delta_{x',x+1}$. To proceed, we must solve two problems.
First, we must find a formulation of the wavefunction that admits an efficient MPS representation. Second, this formulation must also provide a simple matrix product operator (MPO) representation for Eq.~\eqref{eq:hamQ1}. As we shall see, once these two problems are solved, we can use the entire MPO/MPS toolbox with almost no modifications.

\bigskip
\textit{The first quantized MPS.---} The approach that we propose is deceptively simple: we work with a wave-function $\bar\psi$ that encodes only \emph{one} sector of the configuration space, i.e. $\bar\psi =\psi$ for the configurations that satisfy
the constraint $\mathcal{C}$,
\begin{align}
  \mathcal{C} : 0 < x_1 < x_2 < \cdots < x_N < L+1
  \label{eq:R1}
\end{align}
and $\bar\psi = 0$ otherwize. The value of $\psi$ in other sectors can, if necessary, be deduced by permutation. $\bar\psi$ contains the entire information on $\psi$. To see that $\bar\psi$ has a low "entanglement" entropy (entanglement to be now understood in a mathematical sense as a property of the $\bar\psi$ tensor), consider a simple charge density wave at half-filling 
$\bar\psi(x_i)=\delta_{x_1,1}\ldots\delta_{x_p,2p+1}\ldots\delta_{x_N,2N+1}$.
The one particle entanglement entropy of this tensor is $S_1 = 0$ instead of 
the large $S_1 ={\rm{ln}}N$, if one takes all the $N!$ permutations of the $\psi$ tensor into account. To implement $\mathcal{C}$, we take another step: we introduce a change of variable in terms of the inter-particle distance $q_{n} = x_{n} - x_{n-1}$ for $n>1$  with $q_{1} = x_{1}$ and build the technique around $\bar\psi(q_1,\ldots,q_N)$.
The condition $q_i>0$ automatically implements the Pauli principle.
The constraint $\mathcal{C}$ reduces to $x_N = \sum_i q_i \le L$ to prevent the
particles from leaving the box.

\textit{The MPO construction.---} To proceed, we introduce the projector 
$\mathbbm{P}_{m}^n= (\mathbbm{P}_{m}^n)^2$ that acts on variable $q_n$
and projects on a single value $m$: $(\mathbbm{P}_{m}^n)_{qq'}=\delta_{qq'}\delta_{qm}$. We also introduce the
hopping operator $\mathbbm{T}^n$ that acts on the variable $q_n$ and is defined
as $(\mathbbm{T}^n)_{qq'}=\delta_{q',q+1}$ [$\mathbbm{T}^n\bar\psi = \bar\psi(\ldots,q_{n-1},q_n-1,q_{n+1},\ldots$)].
The interaction Hamiltonian $H_V$ takes a simple diagonal form as a sum $H_V=V\sum_{n=2}^N \mathbbm{P}_{1}^n$
which can be cast into an MPO of bond dimension $D=2$:
\begin{align}
 H_V = V
 \begin{bmatrix} 
 0 & \mathbbm{I}
 \end{bmatrix}
 \prod_{n=2}^{N-1}
 \begin{bmatrix*}[l]
 \mathbbm{1} & 0 \\
 \mathbbm{P}_1^n & \mathbbm{1}
 \end{bmatrix*}
 \begin{bmatrix} 
 \mathbbm{1} \\ \mathbbm{P}_1^N
 \end{bmatrix},
\label{eq:H_V}
\end{align}
where $\mathbbm{1}$ is the identity matrix. Second, the hopping term $H_t$
translates into $H_t = -t\sum_{n=1}^{N-1} \mathbbm{T}^{n\dagger}\mathbbm{T}^{n+1}+h.c.$
(moving particle $n$ by one unit on the right increases $q_n$ by one and decreases $q_{n+1}$ by one). It follows that the kinetic terms, $H_t$  has a simple MPO expression with $D=4$,
\begin{align}
 H_{t} = -t
 \begin{bmatrix} 
 0 & \mathbbm{T}^{1} & \mathbbm{T}^{1\dagger} & \mathbbm{1}
 \end{bmatrix}
 \prod_{n=2}^{N-1}
 \begin{bmatrix*}[l]
 \mathbbm{1} & 0 & 0 & 0 \\
 \mathbbm{T}^{n\dagger} & 0 & 0 & 0 \\
 \mathbbm{T}^{n} & 0 & 0 & 0 \\
 0 & \mathbbm{T}^{n} & \mathbbm{T}^{n\dagger} & \mathbbm{1}
 \end{bmatrix*}
 \begin{bmatrix*}[l] 
  \mathbbm{1} \\  \mathbbm{T}^{N\dagger} \\ \mathbbm{T}^{N} \\ 0
 \end{bmatrix*}.
\label{eq:H_K}
\end{align}

It remains to find the MPO $\mathbbm{P}_\mathcal{C}$ that implements
the projection on the configurations that satisfy $\mathcal{C}$.
Formally, it takes the form  $\mathbbm{P}_\mathcal{C} = \sum_{x=1}^L \mathbbm{P}_{x_N = x}$
where the $\mathbbm{P}_{x_\alpha = x}$ projects on the states where $x_\alpha = x$,
\begin{equation}
\mathbbm{P}_{x_\alpha = x} = \sum_{m_1+\ldots +m_\alpha = x} \prod_{n=1}^\alpha \mathbbm{P}_{m_n}^n
\end{equation}
Introducing the $Q\times Q$ lower triangular matrix $\Lambda^n_Q$ defined
as $(\Lambda^n_Q)_{ij} = \mathbbm{P}_{|i-j+1|}^n$ for $i\ge j$ and $0$ otherwise,
we find that 
\begin{equation}
    \mathbbm{P}_{x_\alpha = x} = \left(\prod_{n=1}^\alpha \Lambda^n_{L+1-\alpha}\right)_{i+x-\alpha,i}, \forall i,\ \forall x\ge\alpha
\end{equation}
from which the MPO for $\mathbbm{P}_\mathcal{C}$ follows trivially.
We arrive at the Hamiltonian in MPO form in the $q_n$ coordinates as,  
\begin{align}
H = \mathbbm{P}_\mathcal{C} \left(H_t + H_V\right) \mathbbm{P}_\mathcal{C}.
\label{eq:Q1tV}
\end{align} 
This form of $H$ is, however, unsuitable for numerical calculations because the
rank of $\mathbbm{P}_\mathcal{C}$ scales with 
the system size $D=L+1-N$.
To proceed, we make two \emph{controlled} approximations. First, we work
with $H' = H_t + H_V + \lambda \mathbbm{P}_\mathcal{C}$ which has the
exact same low energy spectrum as $H$ provided the Lagrange multiplier $\lambda$ is large enough. It is easy to quantify any inaccuracy due to this approximation by monitoring any possible finite leakage ($1-\braket{\Psi |\mathbbm{P}_\mathcal{C}| \Psi}\ne 0$) outside of $\mathcal{C}$ (negligible in practice). Second, for a finite density of particles, the allowed values of $q_i$ are in practice very small. For instance at half-filling, 
$\psi(q_1,\ldots,q_n)$ is strongly picked around $q_i=2, \ \forall i$. 
We therefore 
introduce a cut-off $Q_{\rm max}$ in the inter-particle distance and use
$\Lambda^n_{Q_{\rm max}}$ in the calculation of $\mathbbm{P}_\mathcal{C}$. 
In what follows, we have checked that $Q_{\rm max}=10$ is sufficient for our results to be unaffected by this cutoff within our accuracy. We are now in possession of
a low-rank Hamiltonian MPO and can proceed with numerical calculations.
Note that in first quantization, charge conservation is automatic ($N$ is the length of the MPS). This is quite different from the second quantized MPS setting where one must enforce the U$(1)$ charge symmetry (through a particle number block structure implemented on each tensor) to constrain the particle number \cite{Singh2010, Singh2011}.

\begin{figure}[!htb]%left bottom right top
  \includegraphics[width=1.0\linewidth,trim = 0.0in 0.0in 0.0in 0.0in,clip=true]{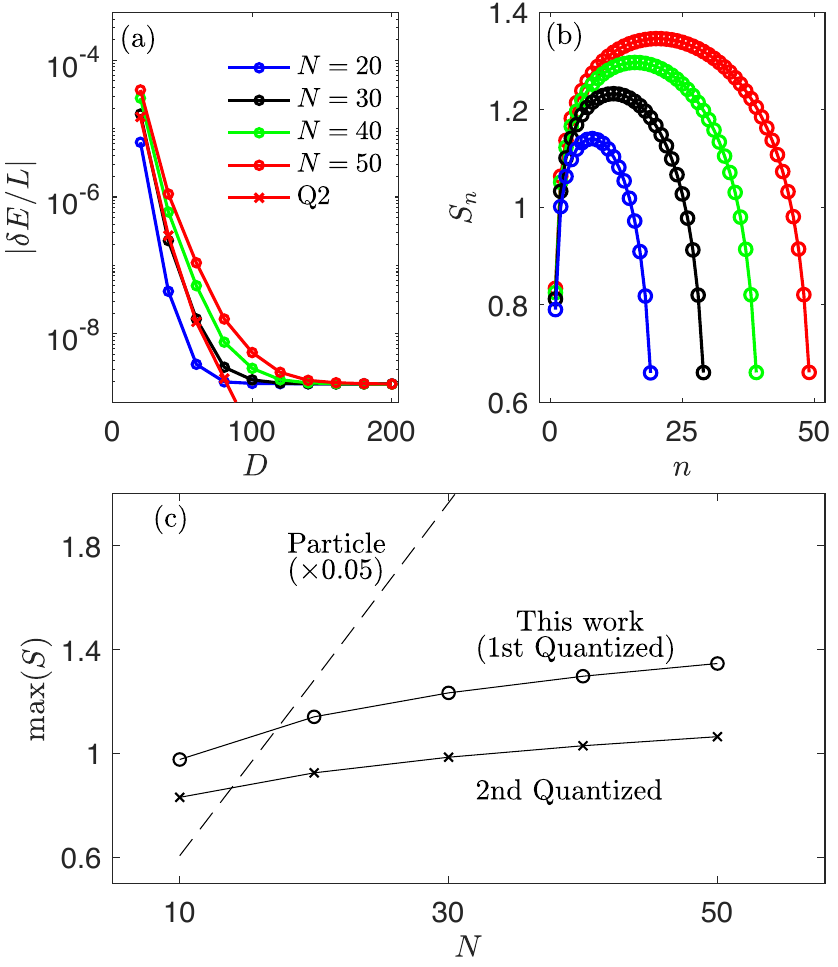}
  \caption{Benchmark: free fermions at half-filling ($L=2N$) with particle numbers $N$ up to $50$.
  (a) The error of energy per site with increasing bond dimensions up to $D_{\rm{max}} = 200$. (Q2: the $2$nd quantized simulation via standard DMRG.)
  (b) The von Neumann entanglement entropy of the first quantized MPS measured at different particle bipartition cut $n$.
  (c) The maximal entanglement entropy, compared between the first quantized MPS ($\circ$), the second quantized MPS (x), and the particle entanglement entropy (--) from the theoretical estimations.  
  }
  \label{fig:GS1}
\end{figure}

\bigskip
\textit{Numerical results.---}
We first compute ground states using the density matrix renormalization group (DMRG) method with single-site update \cite{White1992a, White2005, Hubig2015, Gleis2023} for free and interacting spinless fermions.
We summarize our DMRG results at the non-interacting limit in Fig.\ref{fig:GS1}.
The filling ratio $\rho=N/L$ is fixed at $0.5$, where the wave function is expected to have the most entanglement.

In Fig.\ref{fig:GS1}(a), we plot the error of the ground state energy density, $\delta E/L$. 
As we can see, varying the particle number $N$ from $20$ to $50$, the error systematically goes down below $10^{-8}$ by increasing the MPS bond dimension $D$.
The saturation of the error at large $D$ is due to a small leakage outside of $\mathcal{C}$ with $|1-\braket{\psi|\mathbbm{P}_\mathcal{C}| \psi}| \sim 10^{-7}$.
In comparison with the second quantized DMRG simulation (Q2) at $N=50$, the first quantized MPS exhibits a comparable but slightly slower convergence rate with respect to the bond dimension.

Next, we inspect the bipartite entanglement entropy $S_n$ between the first $n$ particles and the rest $N-n$ one for the first-quantized MPS
(obtained from the singular values of the $n^\text{th}$ tensor of the canonicalized MPS).
$S_n$ is plotted in Fig.~\ref{fig:GS1}(b). It displays a clear deviation from the standard first-quantization one, where theories expect $S_{n} \gtrsim {\rm{ln}}\binom{N}{n}$ for a Slater determinant \cite{Carlen2016, Lemm2017, Radhakrishnan2023a}. Fig.~\ref{fig:GS1}(c) shows the maximal entanglement entropy versus $N$ for the three different approaches 
(naive lower bound of particle entanglement, first quantization and second quantization in real-space).
Particle entanglement entropy suffers an exponential increase, while the first-quantized entropy is comparable (although slightly larger) than its second-quantization counterpart (The particle MPS has been divided by $20$ to keep it on the same scale as the others). In short, these simulations demonstrate that encoding the
wavefunction into $\bar\psi$ provides an efficient MPS approach in first quantization.

\begin{figure}[!htb]%left bottom right top
  \includegraphics[width=1.0\linewidth,trim = 0.0in 0.0in 0.0in 0.0in,clip=true]{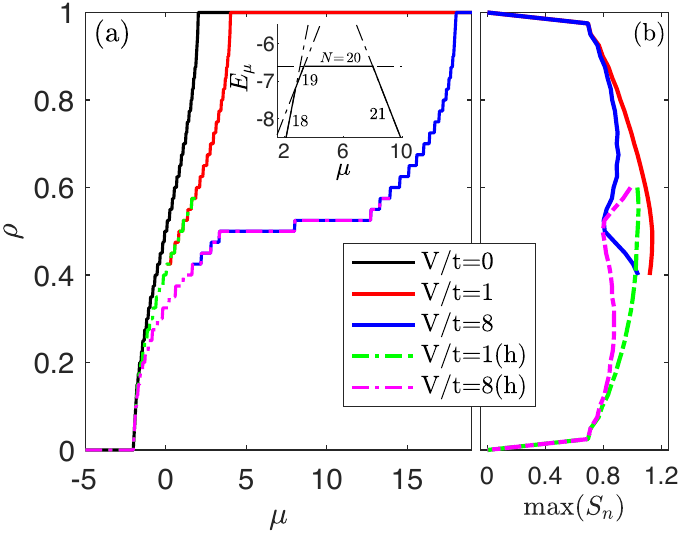}
  \caption{ Ground states for interacting spinless fermions ($V/t=1$ and $8$) with varying particle density at $L=40$. 
  The symbol (h) denotes the hole simulation at low particle density.
  (a) The equation of state via the Maxwell construction.
  Inset: a zoom-in view of the Maxwell construction near half-filling for $V/t=8$.
  (b) The evolution of the maximal entanglement entropy versus $\rho$.
  }
  \label{fig:GS2}
\end{figure}

%%%%%%%%%%%%%%%%%%%%%%%%%%%%%%%%%%%%%%%%%%%%%%%%%%%%%%%%%%%
%%%%%%%%%%%%%%%%%%%%%%%%%%%%%%%%%%%%%%%%%%%%%%%%%%%%%%%%%%%
\textit{Interacting fermions.---}
In DMRG, the $V\ne 0$ case is, in principle, not more difficult than $V=0$. Next, we turn on the interaction $V$, vary the particle number $N$ away from half-filling, and assert that this property remains for first-quantized MPS. From the ground state energy $E(N,V)$ obtained in DMRG, one can easily obtain the equation of state $\rho(\mu)$ \cite{Moreo1991, Dagotto1992a} of the model ($\rho=N/L$: particle density, $\mu$: chemical potential). Indeed adding a term $\mu \sum_l \hat n_l$ is trivial in first quantization, the energy becoming $E_\mu = E(N,V)+\mu N$. Finding the value of $N$ that minimizes the energy for a given $\mu$ simply amounts to computing intersections between subsequent straight lines, see the inset of Fig.~\ref{fig:GS2}(a).

Fig.~\ref{fig:GS2}(a) shows $\rho(\mu)$ for different $V$.
At weak interaction ($V=1$), $\rho$ is a continuous function of $\mu$, similar to the free fermion case (until the band is fully filled).
By contrast, at $V=8$, two distinct plateaux are formed close to the half-filling.
Such discontinuity means that to inject one or two particles at half-filling, substantial energy is needed, which indicates the formation of a charge gap.
This observation concurs with the Bethe ansatz prediction \cite{Yang1966a, Yang1966b}, where at half-filling a gapless Luttinger liquid at small $V$ transitions into a gapped charge density state when $V>2$. Note that at a small particle number, one would need to increase the maximum particle distance $Q_\text{max}$ to account for the dilution. We perform these simulations using the hole Hamiltonian (obtained through the definition of the hole creation operator $b^\dagger_\ell = (-1)^\ell c_l$) with $N_h=L-N$ holes (dashed lines). The two results match perfectly in the $0.4<\rho<0.6$ region where we have used both methods.

The entanglement entropy, see Fig.~\ref{fig:GS2}(b), shows a counter-intuitive behavior.
When a system is interacting, a sum of multiple Slater determinants becomes necessary to describe ground states. Intuitively, this should inevitably lead to an increase of the particle entanglement entropy, as observed numerically \cite{Zozulya2008, Barghathi2017,  Radhakrishnan2023a}. In our framework, however, the interaction $V$ appears to reduce the entanglement entropy close to the half-filling.
For $V=8$, varying the particle density develops a dip of the maximal entanglement entropy, which can be seen from both particle and hole simulations.

\textit{Time evolutions.---} We turn to real-time dynamics and simulate the time evolution of a domain wall motion for free fermions. This problem can also be solved analytically \cite{Antal1999} and is a standard benchmark for numerical studies \cite{Gobert2005, Haegeman2016, Li2022}.

\begin{figure}[!htb]%left bottom right top
  \includegraphics[width=1.0\linewidth,trim = 0.0in 0.0in 0.0in 0.0in,clip=true]{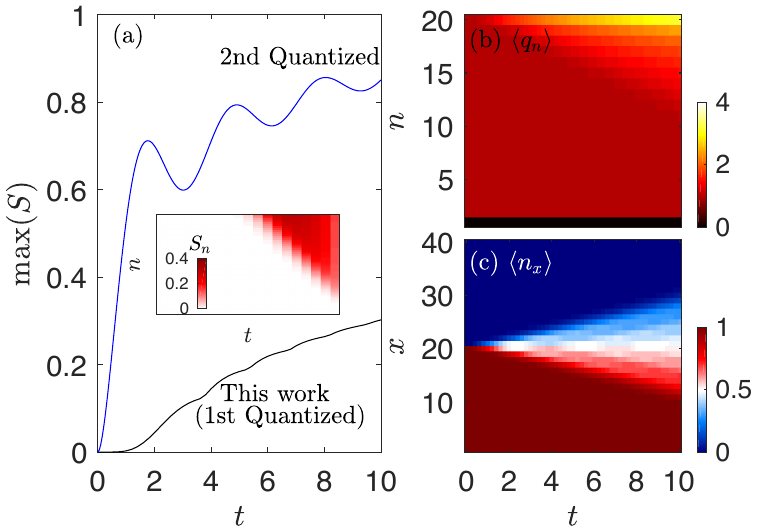}
  \caption{Time evolution of a domain wall at $L=40$. 
  (a) The evolution of maximal entanglement entropy $S_{\rm{vN}}$
  Inset: $S_{\rm{vN}}$ measured on each MPS bond.
  (b) The time profile of inter-particle distance $\braket{q_{n}}$ for $1 \leq n \leq 20$.
  (c) The time profile of occupation number $\braket{n_x}$ on lattice sites for $1 \leq x \leq 40$. 
  }
  \label{fig:TE1}
\end{figure}

The initial state is prepared at half-filling with all electrons put on the left.
The corresponding wavefunction $\psi(q_1,\ldots,q_N)= \prod_i \delta_{q_i,i}$ is a simple product state. We time evolve the MPS via time-dependent variational principle (TDVP) using a finite time step $dt=0.02$ with second-order Trotter decomposition and single-site update \cite{Lubich2015a, Haegeman2016, Yang2020a, Dunnett2021, Ceruti2022, Li2022, Xu2022}.

Fig.~\ref{fig:TE1}(a) shows the growth of entanglement entropy versus time. We observe that 
the first-quantized MPS displays an entanglement growth that is significantly smaller (a factor $\sim 4$, remember that the impact on the bond dimension is exponential) compared to its second-quantized counterpart. 
This difference in behaviour originates from the position of the domain wall at $t=0$. In the second quantization picture, the domain wall corresponds to the middle of the MPS. 
In the first-quantized MPS, the domain wall is situated on the last tensor. Hence, the entanglement growth starts from the boundary of the MPS (see Inset) and takes time to reach the center of the MPS where its growth is not restricted by its position 
(as $S_n<n\log Q_\text{ max}$). 
The measurement of inter-particle distance, $\braket{q_{n}}$, shown in Fig.~\ref{fig:TE1}(b) also agrees with the spread of the entanglement entropy.
Fig.~\ref{fig:TE1}(b) also confirms that the expectation values of $\braket{q_{n}}$ in the simulated time are well below $Q_{\rm{max}}=10$ so that the accuracy of the simulations is not impacted by this cutoff.

Lastly, Fig.~\ref{fig:TE1}(c) shows that one can use the first-quantized MPS to compute quantities in the second quantization picture, here the local occupation number $\braket{n_{x}}$ at lattice site $x$.
An easy calculations show that $\braket{n_{x}} = \sum_{\alpha=1}^N \braket{\mathbbm{P}_{x_\alpha = x}} $. 
Although the local occupation number has become a non-local object in our scheme, the summation can be trivially parallelized.
The time profile of $\braket{n_{x}}$ shows that the initial domain wall develops a light cone propagation, in agreement with the analytical prediction.

\begin{figure}[!htb]%left bottom right top
  \includegraphics[width=1.0\linewidth,trim = 0.0in 0.0in 0.0in 0.0in,clip=true]{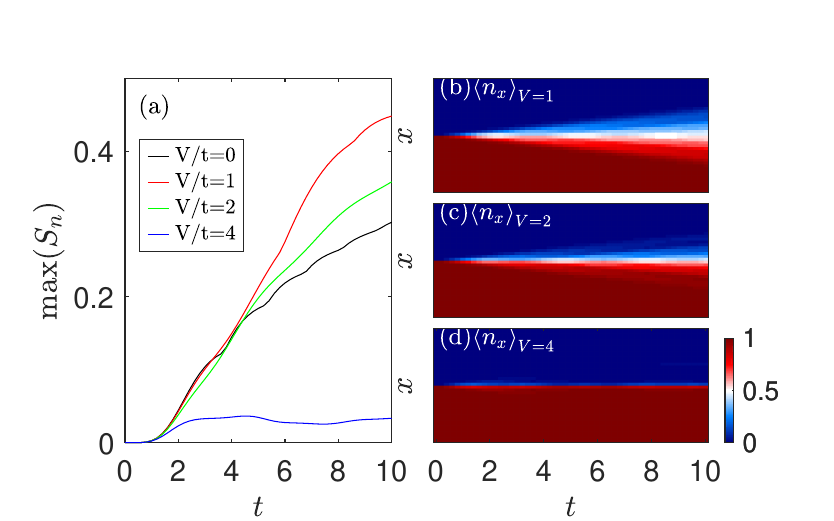}
  \caption{Time evolution of a domain wall for interacting fermions.
  (a) The evolution of maximal entanglement entropy for $V = 0,1,2$ and $4$.
  (b-d) The time profiles of occupation number $\braket{n_x}$.
  }
  \label{fig:TE2}
\end{figure}

Similarly, to the DMRG scenario, we can also perform the time evolution with particle-particle interactions.
The density plots for different $V$ are shown in Fig.~\ref{fig:TE2}(b-d), where the front propagation slows down as $V$ increases.
Such slowdown can be understood quite intuitively, as at $V\rightarrow \infty$, the domain wall configuration becomes an eigenstate of the Hamiltonian.
In Fig.~\ref{fig:TE2}(a), we can see the entanglement entropy grows steadily in time for $V \leq 2$, indicating the delocalization of the fermions from left to right in real-time.
While at $V=4$, there is only local oscillation near the center region, and the growth of entanglement entropy saturates after $t=2$.

\bigskip
\textit{Conclusion.---}
This work reports on a rather intriguing fact: it is possible to formulate a many-body problem within the first quantization framework without significantly raising the entanglement entropy (hence the computational time) with respect to the standard second quantization framework. 
Importantly, the concept of entanglement has two different meanings in the two schemes so that problems that are hard in one scheme may be easy in the other and vice versa. We have identified one first instance where the first-quantization approach is favourable: domain wall propagation. More such problems should be within reach: for instance, we expect that the one-dimensional Wigner cristal with long-range interaction should yield to this approach as the interaction is diagonal in terms of distances (with the further introduction of a minimum cutoff $Q_\text{min}$ for optimization). A better understanding of entanglement in the first quantization scheme (is there the equivalent of an area law?) would be beneficial for future developments.

Much work also remains to be done to explore and optimize the numerical method. In our current scheme, a computational bottleneck is the local physical dimension (set by $Q_\text{max}$) that needs to be sufficiently large. A possible improvement would be to treat the physical index, which labels the particle position, via the so-called quantics representation \cite{Oseledets2009, Khoromskij2011, Ritter2023}, where one maps a real-space grid onto a tuple of qubits.
Such quantic mapping in fact dates back to 1997 \cite{Abrams1997}, originally proposed for quantum computation, however with no experimental realization for more than one fermion so far \cite{Jones2012, Su2021, Chan2023}. The method can also be extended straightforwardly to more complex problems including e.g. spins or bosons.

We end with a remark on research methodology. In the early stage of this project, we have made heavy use of the tensor cross-interpolation (TCI) algorithm that learns an MPS from a callable tensor (a tensor whose value can be computed for any choice of the legs but that is usually to large to be kept in memory) \cite{TCI2022}. For instance, at $V=0$ the exact form of the Slater determinant is known and can be fed to TCI; TCI produces the MPS and more importantly, tells us the value of the bond dimension $D$. Although TCI does not appear anymore in the present work, it is worth remembering its role as a Go/No--go exploration tool in the search for low-rank objects.

\section*{Acknowledgments}
We thank Yuriel N\'u\~nez Fern\'andez for useful discussions.
J.-W.L. acknowledges funding support from the Plan France  2030 ANR-22-PETQ-0007 ``EPIQ''. We thank Yuriel Nu\~nez Fernandez for illuminating discussions.

%\clearpage
\bibliography{MPS_QM1}

\begin{thebibliography}{53}%
\makeatletter
\providecommand \@ifxundefined [1]{%
 \@ifx{#1\undefined}
}%
\providecommand \@ifnum [1]{%
 \ifnum #1\expandafter \@firstoftwo
 \else \expandafter \@secondoftwo
 \fi
}%
\providecommand \@ifx [1]{%
 \ifx #1\expandafter \@firstoftwo
 \else \expandafter \@secondoftwo
 \fi
}%
\providecommand \natexlab [1]{#1}%
\providecommand \enquote  [1]{``#1''}%
\providecommand \bibnamefont  [1]{#1}%
\providecommand \bibfnamefont [1]{#1}%
\providecommand \citenamefont [1]{#1}%
\providecommand \href@noop [0]{\@secondoftwo}%
\providecommand \href [0]{\begingroup \@sanitize@url \@href}%
\providecommand \@href[1]{\@@startlink{#1}\@@href}%
\providecommand \@@href[1]{\endgroup#1\@@endlink}%
\providecommand \@sanitize@url [0]{\catcode `\\12\catcode `\$12\catcode
  `\&12\catcode `\#12\catcode `\^12\catcode `\_12\catcode `\%12\relax}%
\providecommand \@@startlink[1]{}%
\providecommand \@@endlink[0]{}%
\providecommand \url  [0]{\begingroup\@sanitize@url \@url }%
\providecommand \@url [1]{\endgroup\@href {#1}{\urlprefix }}%
\providecommand \urlprefix  [0]{URL }%
\providecommand \Eprint [0]{\href }%
\providecommand \doibase [0]{https://doi.org/}%
\providecommand \selectlanguage [0]{\@gobble}%
\providecommand \bibinfo  [0]{\@secondoftwo}%
\providecommand \bibfield  [0]{\@secondoftwo}%
\providecommand \translation [1]{[#1]}%
\providecommand \BibitemOpen [0]{}%
\providecommand \bibitemStop [0]{}%
\providecommand \bibitemNoStop [0]{.\EOS\space}%
\providecommand \EOS [0]{\spacefactor3000\relax}%
\providecommand \BibitemShut  [1]{\csname bibitem#1\endcsname}%
\let\auto@bib@innerbib\@empty
%</preamble>
\bibitem [{\citenamefont {Vidal}(2003)}]{Vidal2003a}%
  \BibitemOpen
  \bibfield  {author} {\bibinfo {author} {\bibfnamefont {G.}~\bibnamefont
  {Vidal}},\ }\bibfield  {journal} {\bibinfo  {journal} {Physical Review
  Letters}\ }\textbf {\bibinfo {volume} {91}},\ \href
  {https://doi.org/10.1103/physrevlett.91.147902}
  {10.1103/physrevlett.91.147902} (\bibinfo {year} {2003})\BibitemShut
  {NoStop}%
\bibitem [{\citenamefont {Perez-Garcia}\ \emph {et~al.}(2006)\citenamefont
  {Perez-Garcia}, \citenamefont {Verstraete}, \citenamefont {Wolf},\ and\
  \citenamefont {Cirac}}]{PerezGarcia2006}%
  \BibitemOpen
  \bibfield  {author} {\bibinfo {author} {\bibfnamefont {D.}~\bibnamefont
  {Perez-Garcia}}, \bibinfo {author} {\bibfnamefont {F.}~\bibnamefont
  {Verstraete}}, \bibinfo {author} {\bibfnamefont {M.~M.}\ \bibnamefont
  {Wolf}},\ and\ \bibinfo {author} {\bibfnamefont {J.~I.}\ \bibnamefont
  {Cirac}}\ }\href {https://doi.org/10.48550/ARXIV.QUANT-PH/0608197}
  {10.48550/ARXIV.QUANT-PH/0608197} (\bibinfo {year} {2006})\BibitemShut
  {NoStop}%
\bibitem [{\citenamefont {Hastings}(2007)}]{Hastings2007}%
  \BibitemOpen
  \bibfield  {author} {\bibinfo {author} {\bibfnamefont {M.~B.}\ \bibnamefont
  {Hastings}},\ }\href {https://doi.org/10.1088/1742-5468/2007/08/p08024}
  {\bibfield  {journal} {\bibinfo  {journal} {Journal of Statistical Mechanics:
  Theory and Experiment}\ }\textbf {\bibinfo {volume} {2007}},\ \bibinfo
  {pages} {P08024} (\bibinfo {year} {2007})}\BibitemShut {NoStop}%
\bibitem [{\citenamefont {Eisert}\ \emph {et~al.}(2010)\citenamefont {Eisert},
  \citenamefont {Cramer},\ and\ \citenamefont {Plenio}}]{Eisert2010}%
  \BibitemOpen
  \bibfield  {author} {\bibinfo {author} {\bibfnamefont {J.}~\bibnamefont
  {Eisert}}, \bibinfo {author} {\bibfnamefont {M.}~\bibnamefont {Cramer}},\
  and\ \bibinfo {author} {\bibfnamefont {M.~B.}\ \bibnamefont {Plenio}},\
  }\href {https://doi.org/10.1103/revmodphys.82.277} {\bibfield  {journal}
  {\bibinfo  {journal} {Reviews of Modern Physics}\ }\textbf {\bibinfo {volume}
  {82}},\ \bibinfo {pages} {277} (\bibinfo {year} {2010})}\BibitemShut
  {NoStop}%
\bibitem [{\citenamefont {Brand{\~{a}}o}\ and\ \citenamefont
  {Horodecki}(2014)}]{Brandao2014}%
  \BibitemOpen
  \bibfield  {author} {\bibinfo {author} {\bibfnamefont {F.~G. S.~L.}\
  \bibnamefont {Brand{\~{a}}o}}\ and\ \bibinfo {author} {\bibfnamefont
  {M.}~\bibnamefont {Horodecki}},\ }\href
  {https://doi.org/10.1007/s00220-014-2213-8} {\bibfield  {journal} {\bibinfo
  {journal} {Communications in Mathematical Physics}\ }\textbf {\bibinfo
  {volume} {333}},\ \bibinfo {pages} {761} (\bibinfo {year}
  {2014})}\BibitemShut {NoStop}%
\bibitem [{\citenamefont {Verstraete}\ \emph {et~al.}(2008)\citenamefont
  {Verstraete}, \citenamefont {Murg},\ and\ \citenamefont
  {Cirac}}]{Verstraete2008}%
  \BibitemOpen
  \bibfield  {author} {\bibinfo {author} {\bibfnamefont {F.}~\bibnamefont
  {Verstraete}}, \bibinfo {author} {\bibfnamefont {V.}~\bibnamefont {Murg}},\
  and\ \bibinfo {author} {\bibfnamefont {J.}~\bibnamefont {Cirac}},\ }\href
  {https://doi.org/10.1080/14789940801912366} {\bibfield  {journal} {\bibinfo
  {journal} {Advances in Physics}\ }\textbf {\bibinfo {volume} {57}},\ \bibinfo
  {pages} {143} (\bibinfo {year} {2008})}\BibitemShut {NoStop}%
\bibitem [{\citenamefont {Schollwöck}(2011)}]{Schollwoeck2011}%
  \BibitemOpen
  \bibfield  {author} {\bibinfo {author} {\bibfnamefont {U.}~\bibnamefont
  {Schollwöck}},\ }\href {https://doi.org/10.1016/j.aop.2010.09.012}
  {\bibfield  {journal} {\bibinfo  {journal} {Annals of Physics}\ }\textbf
  {\bibinfo {volume} {326}},\ \bibinfo {pages} {96} (\bibinfo {year}
  {2011})}\BibitemShut {NoStop}%
\bibitem [{\citenamefont {Chan}\ and\ \citenamefont {Sharma}(2011)}]{Chan2011}%
  \BibitemOpen
  \bibfield  {author} {\bibinfo {author} {\bibfnamefont {G.~K.-L.}\
  \bibnamefont {Chan}}\ and\ \bibinfo {author} {\bibfnamefont {S.}~\bibnamefont
  {Sharma}},\ }\href {https://doi.org/10.1146/annurev-physchem-032210-103338}
  {\bibfield  {journal} {\bibinfo  {journal} {Annual Review of Physical
  Chemistry}\ }\textbf {\bibinfo {volume} {62}},\ \bibinfo {pages} {465}
  (\bibinfo {year} {2011})}\BibitemShut {NoStop}%
\bibitem [{\citenamefont {Stoudenmire}\ and\ \citenamefont
  {White}(2012)}]{Stoudenmire2012}%
  \BibitemOpen
  \bibfield  {author} {\bibinfo {author} {\bibfnamefont {E.}~\bibnamefont
  {Stoudenmire}}\ and\ \bibinfo {author} {\bibfnamefont {S.~R.}\ \bibnamefont
  {White}},\ }\href {https://doi.org/10.1146/annurev-conmatphys-020911-125018}
  {\bibfield  {journal} {\bibinfo  {journal} {Annual Review of Condensed Matter
  Physics}\ }\textbf {\bibinfo {volume} {3}},\ \bibinfo {pages} {111} (\bibinfo
  {year} {2012})}\BibitemShut {NoStop}%
\bibitem [{\citenamefont {Bridgeman}\ and\ \citenamefont
  {Chubb}(2017)}]{Bridgeman2017}%
  \BibitemOpen
  \bibfield  {author} {\bibinfo {author} {\bibfnamefont {J.~C.}\ \bibnamefont
  {Bridgeman}}\ and\ \bibinfo {author} {\bibfnamefont {C.~T.}\ \bibnamefont
  {Chubb}},\ }\href {https://doi.org/10.1088/1751-8121/aa6dc3} {\bibfield
  {journal} {\bibinfo  {journal} {Journal of Physics A: Mathematical and
  Theoretical}\ }\textbf {\bibinfo {volume} {50}},\ \bibinfo {pages} {223001}
  (\bibinfo {year} {2017})}\BibitemShut {NoStop}%
\bibitem [{\citenamefont {Paeckel}\ \emph {et~al.}(2019)\citenamefont
  {Paeckel}, \citenamefont {Köhler}, \citenamefont {Swoboda}, \citenamefont
  {Manmana}, \citenamefont {Schollwöck},\ and\ \citenamefont
  {Hubig}}]{Paeckel2019a}%
  \BibitemOpen
  \bibfield  {author} {\bibinfo {author} {\bibfnamefont {S.}~\bibnamefont
  {Paeckel}}, \bibinfo {author} {\bibfnamefont {T.}~\bibnamefont {Köhler}},
  \bibinfo {author} {\bibfnamefont {A.}~\bibnamefont {Swoboda}}, \bibinfo
  {author} {\bibfnamefont {S.~R.}\ \bibnamefont {Manmana}}, \bibinfo {author}
  {\bibfnamefont {U.}~\bibnamefont {Schollwöck}},\ and\ \bibinfo {author}
  {\bibfnamefont {C.}~\bibnamefont {Hubig}},\ }\href
  {https://doi.org/10.1016/j.aop.2019.167998} {\bibfield  {journal} {\bibinfo
  {journal} {Annals of Physics}\ }\textbf {\bibinfo {volume} {411}},\ \bibinfo
  {pages} {167998} (\bibinfo {year} {2019})}\BibitemShut {NoStop}%
\bibitem [{\citenamefont {Baiardi}\ and\ \citenamefont
  {Reiher}(2019)}]{Baiardi2019}%
  \BibitemOpen
  \bibfield  {author} {\bibinfo {author} {\bibfnamefont {A.}~\bibnamefont
  {Baiardi}}\ and\ \bibinfo {author} {\bibfnamefont {M.}~\bibnamefont
  {Reiher}},\ }\href {https://doi.org/10.1021/acs.jctc.9b00301} {\bibfield
  {journal} {\bibinfo  {journal} {Journal of Chemical Theory and Computation}\
  }\textbf {\bibinfo {volume} {15}},\ \bibinfo {pages} {3481} (\bibinfo {year}
  {2019})}\BibitemShut {NoStop}%
\bibitem [{\citenamefont {Vanderstraeten}\ \emph {et~al.}(2019)\citenamefont
  {Vanderstraeten}, \citenamefont {Haegeman},\ and\ \citenamefont
  {Verstraete}}]{Vanderstraeten2019}%
  \BibitemOpen
  \bibfield  {author} {\bibinfo {author} {\bibfnamefont {L.}~\bibnamefont
  {Vanderstraeten}}, \bibinfo {author} {\bibfnamefont {J.}~\bibnamefont
  {Haegeman}},\ and\ \bibinfo {author} {\bibfnamefont {F.}~\bibnamefont
  {Verstraete}},\ }\bibfield  {journal} {\bibinfo  {journal} {{SciPost} Physics
  Lecture Notes}\ }\href {https://doi.org/10.21468/scipostphyslectnotes.7}
  {10.21468/scipostphyslectnotes.7} (\bibinfo {year} {2019})\BibitemShut
  {NoStop}%
\bibitem [{\citenamefont {Or{\'{u}}s}(2019)}]{Orus2019}%
  \BibitemOpen
  \bibfield  {author} {\bibinfo {author} {\bibfnamefont {R.}~\bibnamefont
  {Or{\'{u}}s}},\ }\href {https://doi.org/10.1038/s42254-019-0086-7} {\bibfield
   {journal} {\bibinfo  {journal} {Nature Reviews Physics}\ }\textbf {\bibinfo
  {volume} {1}},\ \bibinfo {pages} {538} (\bibinfo {year} {2019})}\BibitemShut
  {NoStop}%
\bibitem [{\citenamefont {Coleman}(1963)}]{Coleman1963}%
  \BibitemOpen
  \bibfield  {author} {\bibinfo {author} {\bibfnamefont {A.~J.}\ \bibnamefont
  {Coleman}},\ }\href {https://doi.org/10.1103/revmodphys.35.668} {\bibfield
  {journal} {\bibinfo  {journal} {Reviews of Modern Physics}\ }\textbf
  {\bibinfo {volume} {35}},\ \bibinfo {pages} {668} (\bibinfo {year}
  {1963})}\BibitemShut {NoStop}%
\bibitem [{\citenamefont {Shi}(2003)}]{Shi2003}%
  \BibitemOpen
  \bibfield  {author} {\bibinfo {author} {\bibfnamefont {Y.}~\bibnamefont
  {Shi}},\ }\href {https://doi.org/10.1103/PhysRevA.67.024301} {\bibfield
  {journal} {\bibinfo  {journal} {Phys. Rev. A}\ }\textbf {\bibinfo {volume}
  {67}},\ \bibinfo {pages} {024301} (\bibinfo {year} {2003})}\BibitemShut
  {NoStop}%
\bibitem [{\citenamefont {Iblisdir}\ \emph {et~al.}(2007)\citenamefont
  {Iblisdir}, \citenamefont {Latorre},\ and\ \citenamefont
  {Or{\'{u}}s}}]{Iblisdir2007}%
  \BibitemOpen
  \bibfield  {author} {\bibinfo {author} {\bibfnamefont {S.}~\bibnamefont
  {Iblisdir}}, \bibinfo {author} {\bibfnamefont {J.~I.}\ \bibnamefont
  {Latorre}},\ and\ \bibinfo {author} {\bibfnamefont {R.}~\bibnamefont
  {Or{\'{u}}s}},\ }\href {https://doi.org/10.1103/physrevlett.98.060402}
  {\bibfield  {journal} {\bibinfo  {journal} {Physical Review Letters}\
  }\textbf {\bibinfo {volume} {98}},\ \bibinfo {pages} {060402} (\bibinfo
  {year} {2007})}\BibitemShut {NoStop}%
\bibitem [{\citenamefont {Haque}\ \emph {et~al.}(2007)\citenamefont {Haque},
  \citenamefont {Zozulya},\ and\ \citenamefont {Schoutens}}]{Haque2007}%
  \BibitemOpen
  \bibfield  {author} {\bibinfo {author} {\bibfnamefont {M.}~\bibnamefont
  {Haque}}, \bibinfo {author} {\bibfnamefont {O.}~\bibnamefont {Zozulya}},\
  and\ \bibinfo {author} {\bibfnamefont {K.}~\bibnamefont {Schoutens}},\ }\href
  {https://doi.org/10.1103/physrevlett.98.060401} {\bibfield  {journal}
  {\bibinfo  {journal} {Physical Review Letters}\ }\textbf {\bibinfo {volume}
  {98}},\ \bibinfo {pages} {060401} (\bibinfo {year} {2007})}\BibitemShut
  {NoStop}%
\bibitem [{\citenamefont {Zozulya}\ \emph {et~al.}(2008)\citenamefont
  {Zozulya}, \citenamefont {Haque},\ and\ \citenamefont
  {Schoutens}}]{Zozulya2008}%
  \BibitemOpen
  \bibfield  {author} {\bibinfo {author} {\bibfnamefont {O.~S.}\ \bibnamefont
  {Zozulya}}, \bibinfo {author} {\bibfnamefont {M.}~\bibnamefont {Haque}},\
  and\ \bibinfo {author} {\bibfnamefont {K.}~\bibnamefont {Schoutens}},\ }\href
  {https://doi.org/10.1103/PhysRevA.78.042326} {\bibfield  {journal} {\bibinfo
  {journal} {Phys. Rev. A}\ }\textbf {\bibinfo {volume} {78}},\ \bibinfo
  {pages} {042326} (\bibinfo {year} {2008})}\BibitemShut {NoStop}%
\bibitem [{\citenamefont {Amico}\ \emph {et~al.}(2008)\citenamefont {Amico},
  \citenamefont {Fazio}, \citenamefont {Osterloh},\ and\ \citenamefont
  {Vedral}}]{Amico2008}%
  \BibitemOpen
  \bibfield  {author} {\bibinfo {author} {\bibfnamefont {L.}~\bibnamefont
  {Amico}}, \bibinfo {author} {\bibfnamefont {R.}~\bibnamefont {Fazio}},
  \bibinfo {author} {\bibfnamefont {A.}~\bibnamefont {Osterloh}},\ and\
  \bibinfo {author} {\bibfnamefont {V.}~\bibnamefont {Vedral}},\ }\href
  {https://doi.org/10.1103/RevModPhys.80.517} {\bibfield  {journal} {\bibinfo
  {journal} {Rev. Mod. Phys.}\ }\textbf {\bibinfo {volume} {80}},\ \bibinfo
  {pages} {517} (\bibinfo {year} {2008})}\BibitemShut {NoStop}%
\bibitem [{\citenamefont {Haque}\ \emph {et~al.}(2009)\citenamefont {Haque},
  \citenamefont {Zozulya},\ and\ \citenamefont {Schoutens}}]{Haque2009}%
  \BibitemOpen
  \bibfield  {author} {\bibinfo {author} {\bibfnamefont {M.}~\bibnamefont
  {Haque}}, \bibinfo {author} {\bibfnamefont {O.~S.}\ \bibnamefont {Zozulya}},\
  and\ \bibinfo {author} {\bibfnamefont {K.}~\bibnamefont {Schoutens}},\ }\href
  {https://doi.org/10.1088/1751-8113/42/50/504012} {\bibfield  {journal}
  {\bibinfo  {journal} {Journal of Physics A: Mathematical and Theoretical}\
  }\textbf {\bibinfo {volume} {42}},\ \bibinfo {pages} {504012} (\bibinfo
  {year} {2009})}\BibitemShut {NoStop}%
\bibitem [{\citenamefont {Killoran}\ \emph {et~al.}(2014)\citenamefont
  {Killoran}, \citenamefont {Cramer},\ and\ \citenamefont
  {Plenio}}]{Killoran2014}%
  \BibitemOpen
  \bibfield  {author} {\bibinfo {author} {\bibfnamefont {N.}~\bibnamefont
  {Killoran}}, \bibinfo {author} {\bibfnamefont {M.}~\bibnamefont {Cramer}},\
  and\ \bibinfo {author} {\bibfnamefont {M.~B.}\ \bibnamefont {Plenio}},\
  }\href {https://doi.org/10.1103/PhysRevLett.112.150501} {\bibfield  {journal}
  {\bibinfo  {journal} {Phys. Rev. Lett.}\ }\textbf {\bibinfo {volume} {112}},\
  \bibinfo {pages} {150501} (\bibinfo {year} {2014})}\BibitemShut {NoStop}%
\bibitem [{\citenamefont {Carlen}\ \emph {et~al.}(2016)\citenamefont {Carlen},
  \citenamefont {Lieb},\ and\ \citenamefont {Reuvers}}]{Carlen2016}%
  \BibitemOpen
  \bibfield  {author} {\bibinfo {author} {\bibfnamefont {E.~A.}\ \bibnamefont
  {Carlen}}, \bibinfo {author} {\bibfnamefont {E.~H.}\ \bibnamefont {Lieb}},\
  and\ \bibinfo {author} {\bibfnamefont {R.}~\bibnamefont {Reuvers}},\ }\href
  {https://doi.org/10.1007/s00220-016-2651-6} {\bibfield  {journal} {\bibinfo
  {journal} {Communications in Mathematical Physics}\ }\textbf {\bibinfo
  {volume} {344}},\ \bibinfo {pages} {655} (\bibinfo {year}
  {2016})}\BibitemShut {NoStop}%
\bibitem [{\citenamefont {Barghathi}\ \emph {et~al.}(2017)\citenamefont
  {Barghathi}, \citenamefont {Casiano-Diaz},\ and\ \citenamefont
  {Maestro}}]{Barghathi2017}%
  \BibitemOpen
  \bibfield  {author} {\bibinfo {author} {\bibfnamefont {H.}~\bibnamefont
  {Barghathi}}, \bibinfo {author} {\bibfnamefont {E.}~\bibnamefont
  {Casiano-Diaz}},\ and\ \bibinfo {author} {\bibfnamefont {A.~D.}\ \bibnamefont
  {Maestro}},\ }\href {https://doi.org/10.1088/1742-5468/aa819a} {\bibfield
  {journal} {\bibinfo  {journal} {Journal of Statistical Mechanics: Theory and
  Experiment}\ }\textbf {\bibinfo {volume} {2017}},\ \bibinfo {pages} {083108}
  (\bibinfo {year} {2017})}\BibitemShut {NoStop}%
\bibitem [{\citenamefont {Lemm}(2017)}]{Lemm2017}%
  \BibitemOpen
  \bibfield  {author} {\bibinfo {author} {\bibfnamefont {M.}~\bibnamefont
  {Lemm}}\ }\href {https://doi.org/10.48550/arXiv.1702.02360}
  {10.48550/arXiv.1702.02360} (\bibinfo {year} {2017})\BibitemShut {NoStop}%
\bibitem [{\citenamefont {Radhakrishnan}\ \emph {et~al.}(2023)\citenamefont
  {Radhakrishnan}, \citenamefont {Thamm}, \citenamefont {Barghathi},
  \citenamefont {Rosenow},\ and\ \citenamefont {Maestro}}]{Radhakrishnan2023a}%
  \BibitemOpen
  \bibfield  {author} {\bibinfo {author} {\bibfnamefont {H.}~\bibnamefont
  {Radhakrishnan}}, \bibinfo {author} {\bibfnamefont {M.}~\bibnamefont
  {Thamm}}, \bibinfo {author} {\bibfnamefont {H.}~\bibnamefont {Barghathi}},
  \bibinfo {author} {\bibfnamefont {B.}~\bibnamefont {Rosenow}},\ and\ \bibinfo
  {author} {\bibfnamefont {A.~D.}\ \bibnamefont {Maestro}},\ }\href
  {https://doi.org/10.1088/1742-5468/ace430} {\bibfield  {journal} {\bibinfo
  {journal} {Journal of Statistical Mechanics: Theory and Experiment}\ }\textbf
  {\bibinfo {volume} {2023}},\ \bibinfo {pages} {083101} (\bibinfo {year}
  {2023})}\BibitemShut {NoStop}%
\bibitem [{\citenamefont {Singh}\ \emph {et~al.}(2010)\citenamefont {Singh},
  \citenamefont {Pfeifer},\ and\ \citenamefont {Vidal}}]{Singh2010}%
  \BibitemOpen
  \bibfield  {author} {\bibinfo {author} {\bibfnamefont {S.}~\bibnamefont
  {Singh}}, \bibinfo {author} {\bibfnamefont {R.~N.~C.}\ \bibnamefont
  {Pfeifer}},\ and\ \bibinfo {author} {\bibfnamefont {G.}~\bibnamefont
  {Vidal}},\ }\href {https://doi.org/10.1103/PhysRevA.82.050301} {\bibfield
  {journal} {\bibinfo  {journal} {Phys. Rev. A}\ }\textbf {\bibinfo {volume}
  {82}},\ \bibinfo {pages} {050301} (\bibinfo {year} {2010})}\BibitemShut
  {NoStop}%
\bibitem [{\citenamefont {Singh}\ \emph {et~al.}(2011)\citenamefont {Singh},
  \citenamefont {Pfeifer},\ and\ \citenamefont {Vidal}}]{Singh2011}%
  \BibitemOpen
  \bibfield  {author} {\bibinfo {author} {\bibfnamefont {S.}~\bibnamefont
  {Singh}}, \bibinfo {author} {\bibfnamefont {R.~N.~C.}\ \bibnamefont
  {Pfeifer}},\ and\ \bibinfo {author} {\bibfnamefont {G.}~\bibnamefont
  {Vidal}},\ }\href {https://doi.org/10.1103/PhysRevB.83.115125} {\bibfield
  {journal} {\bibinfo  {journal} {Phys. Rev. B}\ }\textbf {\bibinfo {volume}
  {83}},\ \bibinfo {pages} {115125} (\bibinfo {year} {2011})}\BibitemShut
  {NoStop}%
\bibitem [{\citenamefont {White}(1992)}]{White1992a}%
  \BibitemOpen
  \bibfield  {author} {\bibinfo {author} {\bibfnamefont {S.~R.}\ \bibnamefont
  {White}},\ }\href {https://doi.org/10.1103/physrevlett.69.2863} {\bibfield
  {journal} {\bibinfo  {journal} {Physical Review Letters}\ }\textbf {\bibinfo
  {volume} {69}},\ \bibinfo {pages} {2863} (\bibinfo {year}
  {1992})}\BibitemShut {NoStop}%
\bibitem [{\citenamefont {White}(2005)}]{White2005}%
  \BibitemOpen
  \bibfield  {author} {\bibinfo {author} {\bibfnamefont {S.~R.}\ \bibnamefont
  {White}},\ }\bibfield  {journal} {\bibinfo  {journal} {Physical Review B}\
  }\textbf {\bibinfo {volume} {72}},\ \href
  {https://doi.org/10.1103/physrevb.72.180403} {10.1103/physrevb.72.180403}
  (\bibinfo {year} {2005})\BibitemShut {NoStop}%
\bibitem [{\citenamefont {Hubig}\ \emph {et~al.}(2015)\citenamefont {Hubig},
  \citenamefont {McCulloch}, \citenamefont {Schollwöck},\ and\ \citenamefont
  {Wolf}}]{Hubig2015}%
  \BibitemOpen
  \bibfield  {author} {\bibinfo {author} {\bibfnamefont {C.}~\bibnamefont
  {Hubig}}, \bibinfo {author} {\bibfnamefont {I.~P.}\ \bibnamefont
  {McCulloch}}, \bibinfo {author} {\bibfnamefont {U.}~\bibnamefont
  {Schollwöck}},\ and\ \bibinfo {author} {\bibfnamefont {F.~A.}\ \bibnamefont
  {Wolf}},\ }\bibfield  {journal} {\bibinfo  {journal} {Physical Review B}\
  }\textbf {\bibinfo {volume} {91}},\ \href
  {https://doi.org/10.1103/physrevb.91.155115} {10.1103/physrevb.91.155115}
  (\bibinfo {year} {2015})\BibitemShut {NoStop}%
\bibitem [{\citenamefont {Gleis}\ \emph {et~al.}(2023)\citenamefont {Gleis},
  \citenamefont {Li},\ and\ \citenamefont {von Delft}}]{Gleis2023}%
  \BibitemOpen
  \bibfield  {author} {\bibinfo {author} {\bibfnamefont {A.}~\bibnamefont
  {Gleis}}, \bibinfo {author} {\bibfnamefont {J.-W.}\ \bibnamefont {Li}},\ and\
  \bibinfo {author} {\bibfnamefont {J.}~\bibnamefont {von Delft}},\ }\href
  {https://doi.org/10.1103/PhysRevLett.130.246402} {\bibfield  {journal}
  {\bibinfo  {journal} {Phys. Rev. Lett.}\ }\textbf {\bibinfo {volume} {130}},\
  \bibinfo {pages} {246402} (\bibinfo {year} {2023})}\BibitemShut {NoStop}%
\bibitem [{\citenamefont {Moreo}\ \emph {et~al.}(1991)\citenamefont {Moreo},
  \citenamefont {Scalapino},\ and\ \citenamefont {Dagotto}}]{Moreo1991}%
  \BibitemOpen
  \bibfield  {author} {\bibinfo {author} {\bibfnamefont {A.}~\bibnamefont
  {Moreo}}, \bibinfo {author} {\bibfnamefont {D.}~\bibnamefont {Scalapino}},\
  and\ \bibinfo {author} {\bibfnamefont {E.}~\bibnamefont {Dagotto}},\ }\href
  {https://doi.org/10.1103/physrevb.43.11442} {\bibfield  {journal} {\bibinfo
  {journal} {Physical Review B}\ }\textbf {\bibinfo {volume} {43}},\ \bibinfo
  {pages} {11442} (\bibinfo {year} {1991})}\BibitemShut {NoStop}%
\bibitem [{\citenamefont {Dagotto}\ \emph {et~al.}(1992)\citenamefont
  {Dagotto}, \citenamefont {Moreo}, \citenamefont {Ortolani}, \citenamefont
  {Poilblanc},\ and\ \citenamefont {Riera}}]{Dagotto1992a}%
  \BibitemOpen
  \bibfield  {author} {\bibinfo {author} {\bibfnamefont {E.}~\bibnamefont
  {Dagotto}}, \bibinfo {author} {\bibfnamefont {A.}~\bibnamefont {Moreo}},
  \bibinfo {author} {\bibfnamefont {F.}~\bibnamefont {Ortolani}}, \bibinfo
  {author} {\bibfnamefont {D.}~\bibnamefont {Poilblanc}},\ and\ \bibinfo
  {author} {\bibfnamefont {J.}~\bibnamefont {Riera}},\ }\href
  {https://doi.org/10.1103/physrevb.45.10741} {\bibfield  {journal} {\bibinfo
  {journal} {Physical Review B}\ }\textbf {\bibinfo {volume} {45}},\ \bibinfo
  {pages} {10741} (\bibinfo {year} {1992})}\BibitemShut {NoStop}%
\bibitem [{\citenamefont {Yang}\ and\ \citenamefont
  {Yang}(1966{\natexlab{a}})}]{Yang1966a}%
  \BibitemOpen
  \bibfield  {author} {\bibinfo {author} {\bibfnamefont {C.~N.}\ \bibnamefont
  {Yang}}\ and\ \bibinfo {author} {\bibfnamefont {C.~P.}\ \bibnamefont
  {Yang}},\ }\href {https://doi.org/10.1103/PhysRev.150.321} {\bibfield
  {journal} {\bibinfo  {journal} {Phys. Rev.}\ }\textbf {\bibinfo {volume}
  {150}},\ \bibinfo {pages} {321} (\bibinfo {year}
  {1966}{\natexlab{a}})}\BibitemShut {NoStop}%
\bibitem [{\citenamefont {Yang}\ and\ \citenamefont
  {Yang}(1966{\natexlab{b}})}]{Yang1966b}%
  \BibitemOpen
  \bibfield  {author} {\bibinfo {author} {\bibfnamefont {C.~N.}\ \bibnamefont
  {Yang}}\ and\ \bibinfo {author} {\bibfnamefont {C.~P.}\ \bibnamefont
  {Yang}},\ }\href {https://doi.org/10.1103/PhysRev.150.327} {\bibfield
  {journal} {\bibinfo  {journal} {Phys. Rev.}\ }\textbf {\bibinfo {volume}
  {150}},\ \bibinfo {pages} {327} (\bibinfo {year}
  {1966}{\natexlab{b}})}\BibitemShut {NoStop}%
\bibitem [{\citenamefont {Antal}\ \emph {et~al.}(1999)\citenamefont {Antal},
  \citenamefont {R\'acz}, \citenamefont {R\'akos},\ and\ \citenamefont
  {Sch\"utz}}]{Antal1999}%
  \BibitemOpen
  \bibfield  {author} {\bibinfo {author} {\bibfnamefont {T.}~\bibnamefont
  {Antal}}, \bibinfo {author} {\bibfnamefont {Z.}~\bibnamefont {R\'acz}},
  \bibinfo {author} {\bibfnamefont {A.}~\bibnamefont {R\'akos}},\ and\ \bibinfo
  {author} {\bibfnamefont {G.~M.}\ \bibnamefont {Sch\"utz}},\ }\href
  {https://doi.org/10.1103/PhysRevE.59.4912} {\bibfield  {journal} {\bibinfo
  {journal} {Phys. Rev. E}\ }\textbf {\bibinfo {volume} {59}},\ \bibinfo
  {pages} {4912} (\bibinfo {year} {1999})}\BibitemShut {NoStop}%
\bibitem [{\citenamefont {Gobert}\ \emph {et~al.}(2005)\citenamefont {Gobert},
  \citenamefont {Kollath}, \citenamefont {Schollwöck},\ and\ \citenamefont
  {Schütz}}]{Gobert2005}%
  \BibitemOpen
  \bibfield  {author} {\bibinfo {author} {\bibfnamefont {D.}~\bibnamefont
  {Gobert}}, \bibinfo {author} {\bibfnamefont {C.}~\bibnamefont {Kollath}},
  \bibinfo {author} {\bibfnamefont {U.}~\bibnamefont {Schollwöck}},\ and\
  \bibinfo {author} {\bibfnamefont {G.}~\bibnamefont {Schütz}},\ }\bibfield
  {journal} {\bibinfo  {journal} {Physical Review E}\ }\textbf {\bibinfo
  {volume} {71}},\ \href {https://doi.org/10.1103/physreve.71.036102}
  {10.1103/physreve.71.036102} (\bibinfo {year} {2005})\BibitemShut {NoStop}%
\bibitem [{\citenamefont {Haegeman}\ \emph {et~al.}(2016)\citenamefont
  {Haegeman}, \citenamefont {Lubich}, \citenamefont {Oseledets}, \citenamefont
  {Vandereycken},\ and\ \citenamefont {Verstraete}}]{Haegeman2016}%
  \BibitemOpen
  \bibfield  {author} {\bibinfo {author} {\bibfnamefont {J.}~\bibnamefont
  {Haegeman}}, \bibinfo {author} {\bibfnamefont {C.}~\bibnamefont {Lubich}},
  \bibinfo {author} {\bibfnamefont {I.}~\bibnamefont {Oseledets}}, \bibinfo
  {author} {\bibfnamefont {B.}~\bibnamefont {Vandereycken}},\ and\ \bibinfo
  {author} {\bibfnamefont {F.}~\bibnamefont {Verstraete}},\ }\bibfield
  {journal} {\bibinfo  {journal} {Physical Review B}\ }\textbf {\bibinfo
  {volume} {94}},\ \href {https://doi.org/10.1103/physrevb.94.165116}
  {10.1103/physrevb.94.165116} (\bibinfo {year} {2016})\BibitemShut {NoStop}%
\bibitem [{\citenamefont {Li}\ \emph {et~al.}(2022)\citenamefont {Li},
  \citenamefont {Gleis},\ and\ \citenamefont {von Delft}}]{Li2022}%
  \BibitemOpen
  \bibfield  {author} {\bibinfo {author} {\bibfnamefont {J.-W.}\ \bibnamefont
  {Li}}, \bibinfo {author} {\bibfnamefont {A.}~\bibnamefont {Gleis}},\ and\
  \bibinfo {author} {\bibfnamefont {J.}~\bibnamefont {von Delft}},\ }\href@noop
  {} {\  (\bibinfo {year} {2022})},\ \Eprint {https://arxiv.org/abs/2208.10972}
  {arXiv:2208.10972 [cond-mat.str-el]} \BibitemShut {NoStop}%
\bibitem [{\citenamefont {Lubich}\ \emph {et~al.}(2015)\citenamefont {Lubich},
  \citenamefont {Oseledets},\ and\ \citenamefont {Vandereycken}}]{Lubich2015a}%
  \BibitemOpen
  \bibfield  {author} {\bibinfo {author} {\bibfnamefont {C.}~\bibnamefont
  {Lubich}}, \bibinfo {author} {\bibfnamefont {I.~V.}\ \bibnamefont
  {Oseledets}},\ and\ \bibinfo {author} {\bibfnamefont {B.}~\bibnamefont
  {Vandereycken}},\ }\href {https://doi.org/10.1137/140976546} {\bibfield
  {journal} {\bibinfo  {journal} {{SIAM} Journal on Numerical Analysis}\
  }\textbf {\bibinfo {volume} {53}},\ \bibinfo {pages} {917} (\bibinfo {year}
  {2015})}\BibitemShut {NoStop}%
\bibitem [{\citenamefont {Yang}\ and\ \citenamefont {White}(2020)}]{Yang2020a}%
  \BibitemOpen
  \bibfield  {author} {\bibinfo {author} {\bibfnamefont {M.}~\bibnamefont
  {Yang}}\ and\ \bibinfo {author} {\bibfnamefont {S.~R.}\ \bibnamefont
  {White}},\ }\bibfield  {journal} {\bibinfo  {journal} {Physical Review B}\
  }\textbf {\bibinfo {volume} {102}},\ \href
  {https://doi.org/10.1103/physrevb.102.094315} {10.1103/physrevb.102.094315}
  (\bibinfo {year} {2020})\BibitemShut {NoStop}%
\bibitem [{\citenamefont {Dunnett}\ and\ \citenamefont
  {Chin}(2021)}]{Dunnett2021}%
  \BibitemOpen
  \bibfield  {author} {\bibinfo {author} {\bibfnamefont {A.~J.}\ \bibnamefont
  {Dunnett}}\ and\ \bibinfo {author} {\bibfnamefont {A.~W.}\ \bibnamefont
  {Chin}},\ }\href {https://doi.org/10.1103/physrevb.104.214302} {\bibfield
  {journal} {\bibinfo  {journal} {Physical Review B}\ }\textbf {\bibinfo
  {volume} {104}},\ \bibinfo {pages} {214302} (\bibinfo {year}
  {2021})}\BibitemShut {NoStop}%
\bibitem [{\citenamefont {Ceruti}\ \emph {et~al.}(2022)\citenamefont {Ceruti},
  \citenamefont {Kusch},\ and\ \citenamefont {Lubich}}]{Ceruti2022}%
  \BibitemOpen
  \bibfield  {author} {\bibinfo {author} {\bibfnamefont {G.}~\bibnamefont
  {Ceruti}}, \bibinfo {author} {\bibfnamefont {J.}~\bibnamefont {Kusch}},\ and\
  \bibinfo {author} {\bibfnamefont {C.}~\bibnamefont {Lubich}},\ }\bibfield
  {journal} {\bibinfo  {journal} {{BIT} Numerical Mathematics}\ }\href
  {https://doi.org/10.1007/s10543-021-00907-7} {10.1007/s10543-021-00907-7}
  (\bibinfo {year} {2022})\BibitemShut {NoStop}%
\bibitem [{\citenamefont {Xu}\ \emph {et~al.}(2022)\citenamefont {Xu},
  \citenamefont {Xie}, \citenamefont {Xie}, \citenamefont {Schollwöck},\ and\
  \citenamefont {Ma}}]{Xu2022}%
  \BibitemOpen
  \bibfield  {author} {\bibinfo {author} {\bibfnamefont {Y.}~\bibnamefont
  {Xu}}, \bibinfo {author} {\bibfnamefont {Z.}~\bibnamefont {Xie}}, \bibinfo
  {author} {\bibfnamefont {X.}~\bibnamefont {Xie}}, \bibinfo {author}
  {\bibfnamefont {U.}~\bibnamefont {Schollwöck}},\ and\ \bibinfo {author}
  {\bibfnamefont {H.}~\bibnamefont {Ma}},\ }\href
  {https://doi.org/10.1021/jacsau.1c00474} {\bibfield  {journal} {\bibinfo
  {journal} {JACS Au}\ }\textbf {\bibinfo {volume} {2}},\ \bibinfo {pages}
  {335} (\bibinfo {year} {2022})}\BibitemShut {NoStop}%
\bibitem [{\citenamefont {Oseledets}(2009)}]{Oseledets2009}%
  \BibitemOpen
  \bibfield  {author} {\bibinfo {author} {\bibfnamefont {I.~V.}\ \bibnamefont
  {Oseledets}},\ }\href {https://doi.org/10.1134/s1064562409050056} {\bibfield
  {journal} {\bibinfo  {journal} {Doklady Mathematics}\ }\textbf {\bibinfo
  {volume} {80}},\ \bibinfo {pages} {653} (\bibinfo {year} {2009})}\BibitemShut
  {NoStop}%
\bibitem [{\citenamefont {Khoromskij}(2011)}]{Khoromskij2011}%
  \BibitemOpen
  \bibfield  {author} {\bibinfo {author} {\bibfnamefont {B.~N.}\ \bibnamefont
  {Khoromskij}},\ }\href {https://doi.org/10.1007/s00365-011-9131-1} {\bibfield
   {journal} {\bibinfo  {journal} {Constructive Approximation}\ }\textbf
  {\bibinfo {volume} {34}},\ \bibinfo {pages} {257} (\bibinfo {year}
  {2011})}\BibitemShut {NoStop}%
\bibitem [{\citenamefont {Ritter}\ \emph {et~al.}(2023)\citenamefont {Ritter},
  \citenamefont {Fernández}, \citenamefont {Wallerberger}, \citenamefont {von
  Delft}, \citenamefont {Shinaoka},\ and\ \citenamefont
  {Waintal}}]{Ritter2023}%
  \BibitemOpen
  \bibfield  {author} {\bibinfo {author} {\bibfnamefont {M.~K.}\ \bibnamefont
  {Ritter}}, \bibinfo {author} {\bibfnamefont {Y.~N.}\ \bibnamefont
  {Fernández}}, \bibinfo {author} {\bibfnamefont {M.}~\bibnamefont
  {Wallerberger}}, \bibinfo {author} {\bibfnamefont {J.}~\bibnamefont {von
  Delft}}, \bibinfo {author} {\bibfnamefont {H.}~\bibnamefont {Shinaoka}},\
  and\ \bibinfo {author} {\bibfnamefont {X.}~\bibnamefont {Waintal}}\ }\href
  {https://doi.org/10.48550/ARXIV.2303.11819} {10.48550/ARXIV.2303.11819}
  (\bibinfo {year} {2023})\BibitemShut {NoStop}%
\bibitem [{\citenamefont {Abrams}\ and\ \citenamefont
  {Lloyd}(1997)}]{Abrams1997}%
  \BibitemOpen
  \bibfield  {author} {\bibinfo {author} {\bibfnamefont {D.~S.}\ \bibnamefont
  {Abrams}}\ and\ \bibinfo {author} {\bibfnamefont {S.}~\bibnamefont {Lloyd}},\
  }\href {https://doi.org/10.1103/PhysRevLett.79.2586} {\bibfield  {journal}
  {\bibinfo  {journal} {Phys. Rev. Lett.}\ }\textbf {\bibinfo {volume} {79}},\
  \bibinfo {pages} {2586} (\bibinfo {year} {1997})}\BibitemShut {NoStop}%
\bibitem [{\citenamefont {Jones}\ \emph {et~al.}(2012)\citenamefont {Jones},
  \citenamefont {Whitfield}, \citenamefont {McMahon}, \citenamefont {Yung},
  \citenamefont {Meter}, \citenamefont {Aspuru-Guzik},\ and\ \citenamefont
  {Yamamoto}}]{Jones2012}%
  \BibitemOpen
  \bibfield  {author} {\bibinfo {author} {\bibfnamefont {N.~C.}\ \bibnamefont
  {Jones}}, \bibinfo {author} {\bibfnamefont {J.~D.}\ \bibnamefont
  {Whitfield}}, \bibinfo {author} {\bibfnamefont {P.~L.}\ \bibnamefont
  {McMahon}}, \bibinfo {author} {\bibfnamefont {M.-H.}\ \bibnamefont {Yung}},
  \bibinfo {author} {\bibfnamefont {R.~V.}\ \bibnamefont {Meter}}, \bibinfo
  {author} {\bibfnamefont {A.}~\bibnamefont {Aspuru-Guzik}},\ and\ \bibinfo
  {author} {\bibfnamefont {Y.}~\bibnamefont {Yamamoto}},\ }\href
  {https://doi.org/10.1088/1367-2630/14/11/115023} {\bibfield  {journal}
  {\bibinfo  {journal} {New Journal of Physics}\ }\textbf {\bibinfo {volume}
  {14}},\ \bibinfo {pages} {115023} (\bibinfo {year} {2012})}\BibitemShut
  {NoStop}%
\bibitem [{\citenamefont {Su}\ \emph {et~al.}(2021)\citenamefont {Su},
  \citenamefont {Berry}, \citenamefont {Wiebe}, \citenamefont {Rubin},\ and\
  \citenamefont {Babbush}}]{Su2021}%
  \BibitemOpen
  \bibfield  {author} {\bibinfo {author} {\bibfnamefont {Y.}~\bibnamefont
  {Su}}, \bibinfo {author} {\bibfnamefont {D.~W.}\ \bibnamefont {Berry}},
  \bibinfo {author} {\bibfnamefont {N.}~\bibnamefont {Wiebe}}, \bibinfo
  {author} {\bibfnamefont {N.}~\bibnamefont {Rubin}},\ and\ \bibinfo {author}
  {\bibfnamefont {R.}~\bibnamefont {Babbush}},\ }\href
  {https://doi.org/10.1103/PRXQuantum.2.040332} {\bibfield  {journal} {\bibinfo
   {journal} {PRX Quantum}\ }\textbf {\bibinfo {volume} {2}},\ \bibinfo {pages}
  {040332} (\bibinfo {year} {2021})}\BibitemShut {NoStop}%
\bibitem [{\citenamefont {Chan}\ \emph {et~al.}(2023)\citenamefont {Chan},
  \citenamefont {Meister}, \citenamefont {Jones}, \citenamefont {Tew},\ and\
  \citenamefont {Benjamin}}]{Chan2023}%
  \BibitemOpen
  \bibfield  {author} {\bibinfo {author} {\bibfnamefont {H.~H.~S.}\
  \bibnamefont {Chan}}, \bibinfo {author} {\bibfnamefont {R.}~\bibnamefont
  {Meister}}, \bibinfo {author} {\bibfnamefont {T.}~\bibnamefont {Jones}},
  \bibinfo {author} {\bibfnamefont {D.~P.}\ \bibnamefont {Tew}},\ and\ \bibinfo
  {author} {\bibfnamefont {S.~C.}\ \bibnamefont {Benjamin}},\ }\bibfield
  {journal} {\bibinfo  {journal} {Science Advances}\ }\textbf {\bibinfo
  {volume} {9}},\ \href {https://doi.org/10.1126/sciadv.abo7484}
  {10.1126/sciadv.abo7484} (\bibinfo {year} {2023})\BibitemShut {NoStop}%
\bibitem [{\citenamefont {N\'u\~nez Fern\'andez}\ \emph
  {et~al.}(2022)\citenamefont {N\'u\~nez Fern\'andez}, \citenamefont {Jeannin},
  \citenamefont {Dumitrescu}, \citenamefont {Kloss}, \citenamefont {Kaye},
  \citenamefont {Parcollet},\ and\ \citenamefont {Waintal}}]{TCI2022}%
  \BibitemOpen
  \bibfield  {author} {\bibinfo {author} {\bibfnamefont {Y.}~\bibnamefont
  {N\'u\~nez Fern\'andez}}, \bibinfo {author} {\bibfnamefont {M.}~\bibnamefont
  {Jeannin}}, \bibinfo {author} {\bibfnamefont {P.~T.}\ \bibnamefont
  {Dumitrescu}}, \bibinfo {author} {\bibfnamefont {T.}~\bibnamefont {Kloss}},
  \bibinfo {author} {\bibfnamefont {J.}~\bibnamefont {Kaye}}, \bibinfo {author}
  {\bibfnamefont {O.}~\bibnamefont {Parcollet}},\ and\ \bibinfo {author}
  {\bibfnamefont {X.}~\bibnamefont {Waintal}},\ }\href
  {https://doi.org/10.1103/PhysRevX.12.041018} {\bibfield  {journal} {\bibinfo
  {journal} {Phys. Rev. X}\ }\textbf {\bibinfo {volume} {12}},\ \bibinfo
  {pages} {041018} (\bibinfo {year} {2022})}\BibitemShut {NoStop}%
\end{thebibliography}%
\end{document}